\documentstyle[11pt,newpasp,twoside,epsf]{article}
\def\edcomment#1{\iffalse\marginpar{\raggedright\sl#1\/}\else\relax\fi}
\reversemarginpar

\begin{document}
\title{Large--Scale Structure Studies with Clusters of Galaxies}
 \author{R. C. Nichol}
\affil{Dept. of Physics, Carnegie Mellon University, 5000 Forbes Ave., Pittsburgh, PA-15232}

\begin{abstract}
I present here a review of Large-Scale Structure (LSS) studies using clusters
of galaxies. First, I re-evaluate the `pros' and `cons' of using clusters for
such studies, especially in this era of large galaxy redshift surveys.
Secondly, I provide an historical review of the Cluster Correlation Function
and show that the latest measurements of $\xi_{cc}$ from Abell and X--ray
catalogs are in excellent agreement. Thirdly, I review the latest measurements
of the power spectrum of clusters which provide strong constraints on
the cosmological parameters ({\it e.g.} $\Omega_m$) and models of structure
formation. Moreover, I highlight the recent discovery of ``Baryon Wiggles'' in
the local cluster p(k) which is in perfect agreement with the recent CMB
data. Lastly, I examine recent advances in the measurement of the X--ray
Cluster Luminosity Function and emphasize the importance of accurately
determining the selection function of future cluster surveys.
\end{abstract}

\section{Introduction}

I present here an incomplete and personal review of Large-Scale Structure
(LSS) studies using clusters of galaxies. For a more complete overview of
clusters as cosmological probes, the reader is referred to several excellent
recent reviews by Biviano (2000), Schindler (2001a), Borgani \& Guzzo (2001),
Guzzo (2001) and Haiman, Mohr \& Holder (2001).

\subsection{Pros and Cons of Using Clusters for LSS Studies}

As cosmologists, we want to study and understand the distribution of matter in
the Universe as a function of space and time {\it i.e.}  mass(x,y,z,t). This
will allow us to constrain cosmological parameters, models of structure
formation as well as understanding the physical processes that control the
distribution of baryons. Ultimately, we want to directly compare our
observations of the Universe to cosmological simulations {\it e.g.} the
VIRGO-GIF project (Kauffmann et al. 1999). This is hard to do since we need to
know how the light we observe, traces the underlying matter we simulate.

Historically, clusters of galaxies have been used for this task because: {\it
a)} They survey large volumes of space thus producing a fair sample of the
Universe; {\it b)} Clusters can be seen over a large range in redshift, thus
probing cosmological evolution; {\it c)} They maintain the imprint of the
initial conditions so simple analytical relationships can be used to explain
their properties ({\it e.g.} Jenkins et al. 2001); {\it d)} Clusters are the
largest gravitationally bound objects in the Universe, so we can weigh the
Cosmos using them; {\it e)} Clusters are full of hot baryons, so we can study
how mass follows light on cluster scales.

However, clusters do have their problems. First, they live in the tail of the
underlying mass distribution and thus may be highly biased tracers of this
distribution. This has been a major concern for a long time (Kaiser 1986), but
recent observational and theoretical work (see Miller et al. 2001a, Narayanan
et al. 2000) shows that a simple linear biasing model ($b=1.5$), between
normal galaxies and clusters, does work well over an important range of scales
($0.03\le k \le 0.15h^{-1}\,{\rm Mpc^{-1}}$.  As an aside, I note that because
of this fact, clusters may be an excellent probe of the gaussianity of the
Universe since our models and simulations of the Universe assume gaussianity
at the beginning ({\it e.g.}  Chiu, Ostriker, \& Strauss 1998; Robinson,
Gawiser \& Silk 1999; Matarrese, Verde \& Jimenez 2000; Kerscher et al. 2001).
Kerscher et al. (2001) has already found statistical evidence for
non-gaussianity based on the REFLEX sample of X--ray clusters, although Verde
et al.  (2001) highlight that the CMB and high redshift galaxies maybe a
better way to studying this issue.

Another concern with clusters is the assumption of hydrostatic equilbrium (for
the gas) and the virial theorem (for the galaxies). For example, in the recent
hydrodynamical/N-body simulations of Ricker \& Sarazin (2001), an off-axis
merger of two systems (mass ratio of 3:1) can produce large-scale turbulent
motions with eddys up to several 100 kpcs.  Even after nearly a Hubble time,
these motions persist as subsonic turbulence in the cluster cores, providing
5-10\% of the support against gravity.  Roettiger, Burns \& Loken (1996) and
Ritchie \& Thomas (2001) also found that major merger events can knock the
cluster out of hydrostatic equilibrium for several Gyrs (see Schindler et
al. 2001b for a recent review of cluster simulations).

To assess the importance of such mergers on the whole cluster population, I
show in Fig. 1 the expected fraction of clusters ($M\ge10^{14}\,M_{\odot}$ or
$L_{\rm x}(0.5-2.0) > 10^{43}\,{\rm erg/s}$) that have experienced a merger in
the last 2.5Gyrs. This figure shows that as we push to higher redshifts the
vast majority of clusters should show significant evidence for a recent merger
that may severely effect their physical state (see Mathiesen \& Evrard
2001). Fig. 1 is in agreement with the recent {\it Chandra} observations
reported in Henry (2001) where 75\% of $z>0.75$ clusters showed significant
substructure. Therefore, if we wish to continue to use clusters as
cosmological tracers, we must understand such effects in greater detail.


\begin{figure}[t]
\epsfxsize=3.8in 
\centerline{\epsfbox{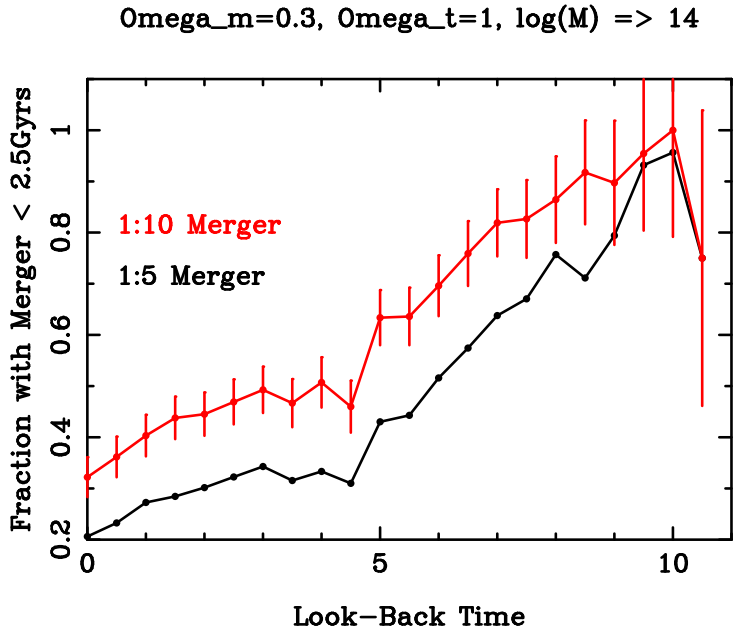}}
\caption{The fraction of $M\ge10^{14}\,M_{\odot}$ clusters that have
experienced a major merger in the last 2.5Gyrs versus look-back time. The red
points are for a 10:1 merger, while the black points are for a 5:1. These data
were kindly provided by Joanne Cohn based on simulations similar to the work
presented in Cohn, Bagla \& White (2000). The simulations use
$\Omega_{m}=0.3$, $\Omega_{\Lambda}=0.7$ and $\sigma_8=0.9$. The error bars
are $\sqrt{N}$ of clusters detected in the simulations. For clarity, they are
not shown for the 5:1 simulations.
\label{merge}
}
\end{figure}

In light of these concerns, I challenged the audience at the conference to
discuss the following question over the coffee break: ``In this era of galaxy
redshift, lensing and photometric surveys$\footnote{Surveys like the SDSS,
2dF, LSST, SNAP, POI, DEEP \& VIRMOS}$, is there still a need for cluster
surveys of the Universe?''. For example, in Fig. 2, I show initial results
from the Sloan Digital Sky Survey (SDSS) Luminous Red Galaxy (LRG) survey
which is designed to create a pseudo--volume limited sample of 100,000 giant
ellipticals out to $z\sim0.4$ (although the sample does reach out to
$z\sim0.6$). This survey will sample the LSS on Gigaparsec scales --
comparable with the largest cluster surveys -- well beyond the turn--over in
the matter power spectrum (see Section 3). The LRGs are still biased tracers
but there is compelling evidence that we can fully understand their selection
function (see Eisenstein et al. 2001 for details).

\begin{figure}[t]
\epsfxsize=4.5in 
\centerline{\epsfbox{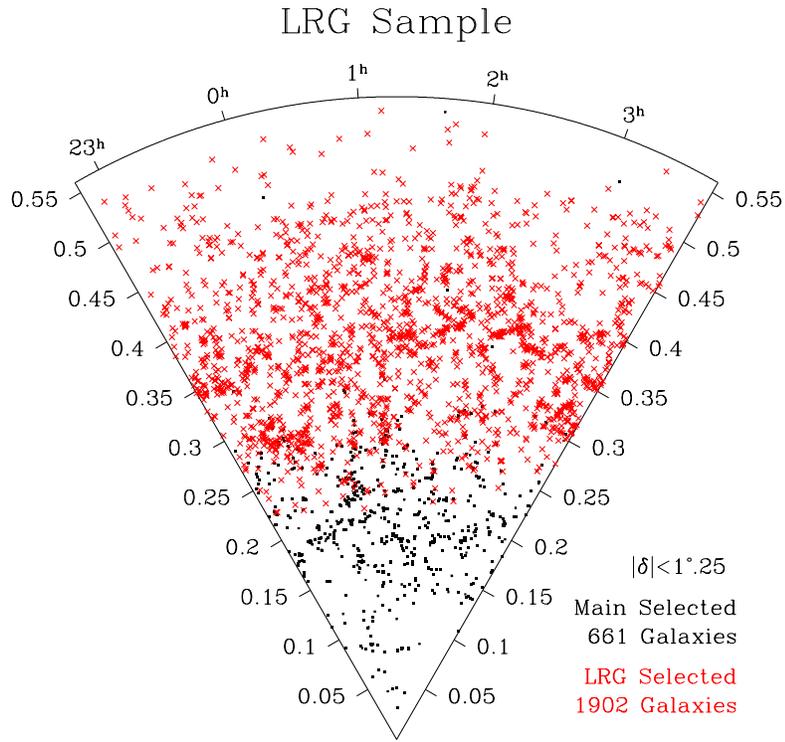}}
\caption{Wedge plot of a small fraction of SDSS spectroscopic data collected
to date. This illustrates the power of using the SDSS--LRG sample for studying
the LSS. Figure kindly provided by Daniel
Eisenstein.
\label{lrg}
}
\end{figure}

Returning to my question, I will emphasize that clusters are still the best
way to: {\it a)} Constrain $\sigma_8$ (see Viana \& Liddle 1999); {\it b)}
Determine $\Omega_{m}$ (see Sections 3 \& 4), and; {\it c)} Study galaxy
evolution in dense regions.

\section{Cluster Correlation Function}

The Cluster Correlation Function ($\xi_{cc}=(\frac{r}{r_o})^{-\gamma}$ where
$r_o$ is the correlation length defined as $\xi_{cc}(r_o)=1$) has had a long
and controversial history. In the 1980's, it was one of the main constraints
on CDM models of structure formation (see White et al. 1987). The main
controversy surrounding measurements of $\xi_{cc}$ has been the effects of
optical projection effects on the Abell catalog {\it i.e.}  the contamination
of the 2D richness of clusters because of the field or a nearby clusters
(Sutherland et al. 1988) and/or the super-position of two groups along the
line--of--sight to create a ``Phantom Cluster''. The hypothesis has been that
these projection effects have artificially boosted the amplitude of Abell
measurements of $\xi_{cc}$ thus making it inconsistent with CDM models
(Sutherland et al. 1988; Dekel et al. 1989; Efstathiou et al. 1992). There are
many advocates of this hypothesis ({\it e.g.} Dalton et al. 1992; Nichol et
al. 1992 \& 1994; Romer et al. 1994), but there are also many defenders of the
Abell catalog (Postman et al. 1992; Miller et al. 1999). Over the last decade,
such concerns have driven the creation of new optical and X--ray cluster
surveys {\it e.g.} the EDCC (Lumdsen et al. 1992), APM (Dalton et al. 1992),
DPOSS (Gal et al. 2000) and REFLEX surveys (Collins et al. 2000).

In Fig. 3, I show an historical overview of $\xi_{cc}$ measurements over the
last 20 years. There are two interesting points to make about this figure:
First, the correlation length of Abell $\xi_{cc}$'s have decreased with
time. Secondly, X--ray measurements of $r_o$ have increased with time. These
two trends can be understood via the relative space density of clusters in
these different surveys. In the case of the Abell survey, the volume surveyed
by the different samples of Abell clusters has not increased much since the
first measurements, but the number of systems used in the measurement of
$\xi_{cc}$ has steadily increased from $\sim100$ to over $600$. Thus the space
density of Abell clusters used in determining $\xi_{cc}$ has decreased over
time. The opposite is true for the X--ray samples; the number of clusters used
has increased, but so has the volume sampled, thus the space density has
slightly decreased over time (see Croft et al 1997).

\begin{figure}[t]  
\epsfxsize=4.2in 
\centerline{\epsfbox{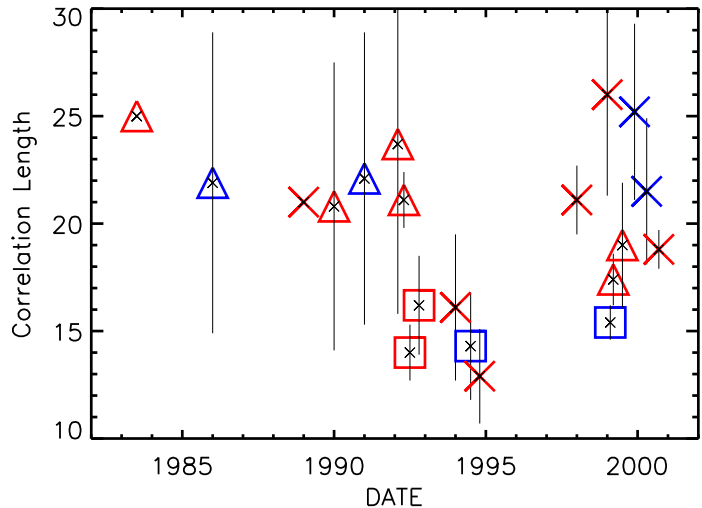}}
\caption{The Correlation Length ($r_o$ in Mpc) of $\xi_{cc}$, measured by
various authors, as a function of time. Triangles are Abell, squares are
optical non-Abell samples and crosses are X--ray samples. Red points have a
slope of the correlation function between $1.7 \le \gamma \le 2.0$, while
blue points are outside this range. Please note that I have not scaled these
measurements by the space density of the cluster sample used in the
measurement. I took the $r_o$ value and error quoted in the abstract of the
papers and, in the case of multiple fits to the data, I took the $r_o$ from
the largest sample. Data from Ling et al. 1986; West \& van der Bergh (1991);
Peacock \& West 1992; Miller et al. 1999; Lahav et al. 1989; Romer et
al. 1994; Borgani et al. 1999a; Abadi et al. 1998; Lee et al. 1999; Moscardini
et al. 2000; Collins et al. 2000; Dalton et al. 1992; Nichol et al. 1992 \&
1994; Postman et al. 1992; Huchra et al. 1990; Bahcall \& Soneria 1983.
\label{xi}
}
\end{figure}

If we focus on recent measurements with similar slopes ($1.7\le \gamma \le
2.0$), then the agreement between present measurements of $\xi_{cc}$ is very
good. This is illustrated in Fig. 4 where I show the $\xi_{cc}$ from nearly
1000 clusters (both Abell and X--ray). The great debate over the value of the
correlation length of $\xi_{cc}$ is probably over especially given that the
pi-sigma diagrams shown in Fig. 5 highlights that the clustering in both the
Abell and X--ray surveys is now nearly isotropic. This removes one of the main
arguments for projection effects in the Abell catalog and is probably due to
the larger samples now available averaging over the effect of LSS and/or the
effect of a single supercluster. We look forward to new
$\xi_{cc}$ measurements from large, homogeneous samples of clusters taken from
the SDSS, REFLEX and 2dF surveys.

\begin{figure}[t]  
\epsfxsize=3.8in 
\centerline{\epsfbox{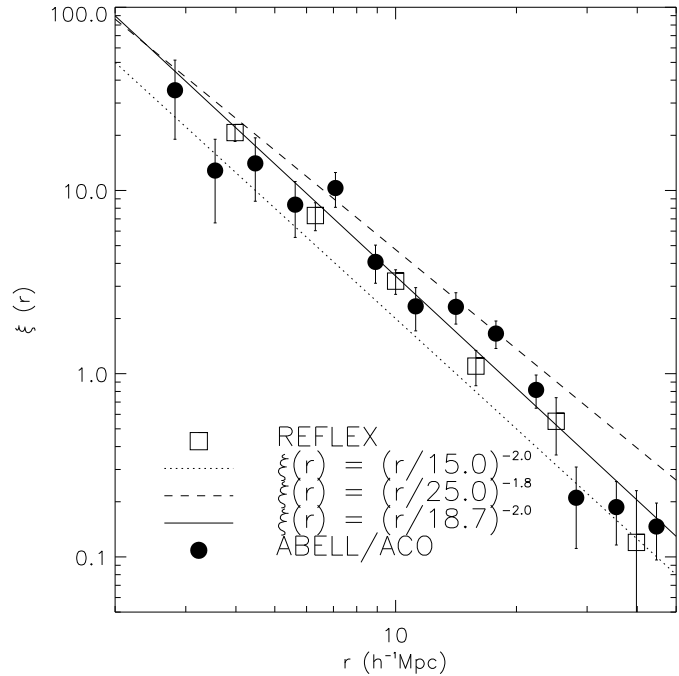}}
\caption{
The correlation function for the Abell/ACO sample of Miller et
al. (1999) and from the REFLEX sample of Collins et al. (2000). No correction
has been made for the space densities of the surveys. Data kindly provided by
Chris Collins and Chris Miller.}
\end{figure}

\begin{figure}[t] 
\vspace{-0.8in} 
\centerline{\epsfxsize=3.0in\epsfbox{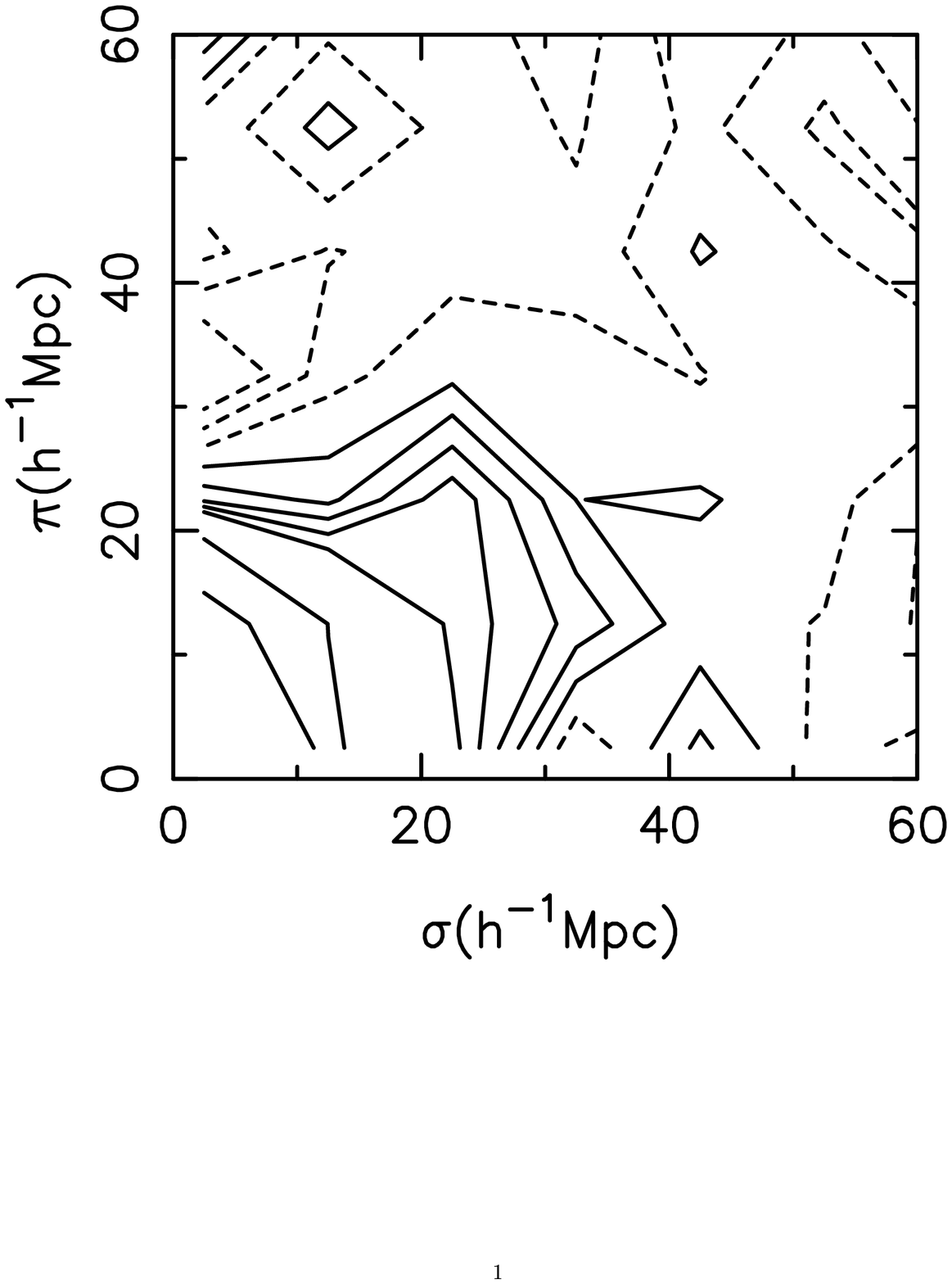}\epsfxsize=2.4in\epsfbox{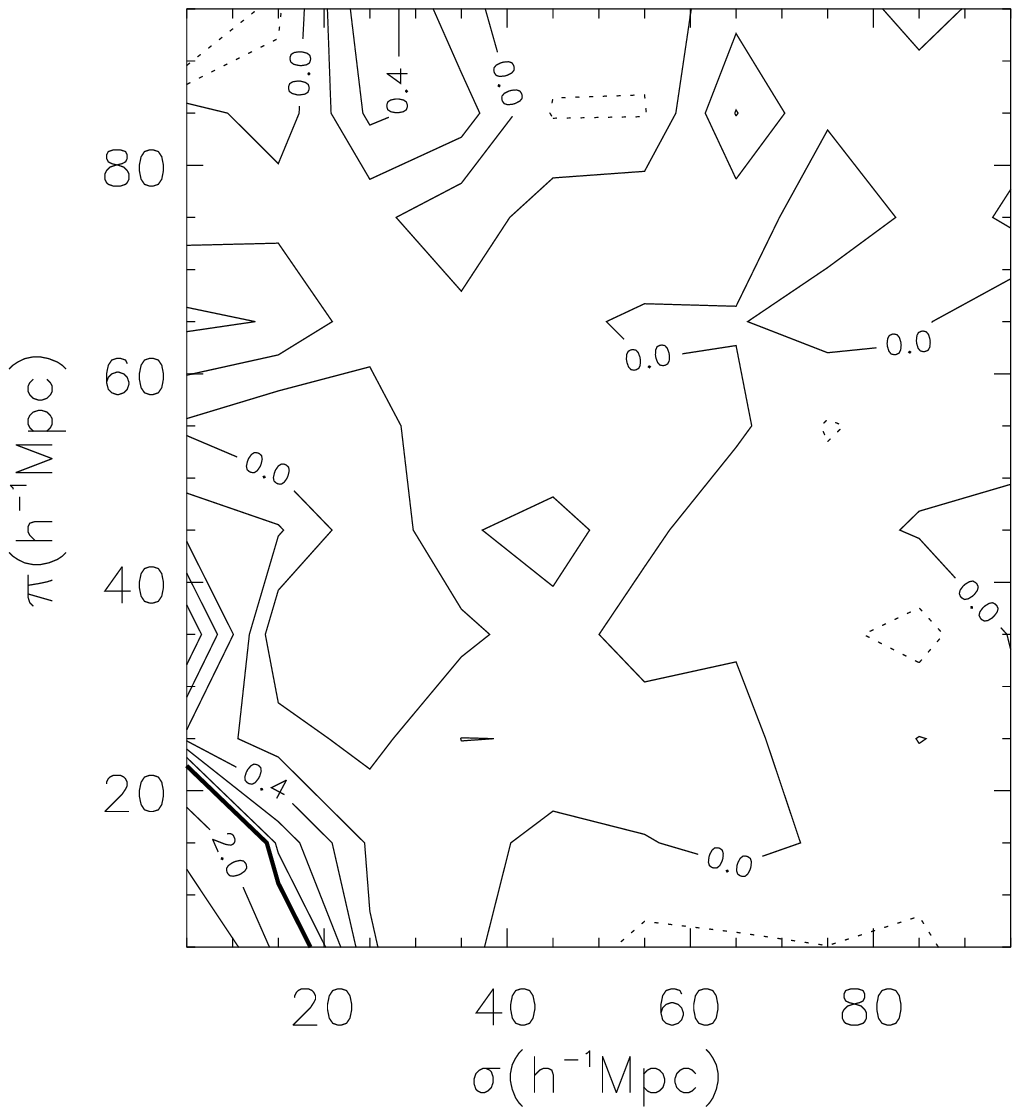}}
\caption{The $\pi$--$\sigma$ diagrams for the two surveys presented in Fig
4. This shows the radial ($\sigma$) and transverse ($\pi$) components of the
correlation function. The left--hand plot is for the REFLEX sample, while the
right--hand side is from the Abell/ACO sample of Miller et al. (1999). These
figures were kindly provided by Chris Collins and Chris Miller.
}
\end{figure}

\section{Cluster Power Spectrum}

In the last ten years, many authors have switched from measuring correlation
functions to measurements of the cluster power spectrum since it allows for a
more direct comparison with the theoretical models. For example, increasing
$\Omega_m h^2$ pushes the epoch of matter-radiation equality back in time
and moves the peak in the p(k) to low k values (Tegmark 1999
$\footnote{See the
CMB movies on Max Tegmark's homepage which illustrate this effect
{\it i.e.} http://www.hep.upenn.edu/\~\,max/}$). 
Therefore, the
detection of the turn--over in the local power spectrum (p(k)) of matter
provides a powerful constrain on both the matter density of the Universe and
models of structure formation ({\it e.g.} CDM).

As yet, there has been no definitive detection of a turn--over (or peak) in
the local matter p(k), but we are getting close (Szalay et al. 2001; Miller et
al. 2001a; Percival et al. 2001; Huterer et al. 2001) as new surveys of the
local Universe are approaching volumes than demand a turn--over to satisfy the
COBE measurements at very low k. This lack of a detection naturally drives
estimates of $\Omega_m$ from these p(k)'s towards low values {\it e.g.} Miller
et al. (2001a) recently estimated $\Omega_m\,h^2 = 0.12^{+0.03}_{-0.02}$ using
a large sample of clusters to probe Gigaparsec scales. Similarly, Schuecker et
al. (2001) presented the p(k) from the REFLEX sample and finds it favors
low--density CDM models as the peak in the p(k) is yet undetected.

\begin{figure}[t]  
\epsfxsize=5.5in 
\centerline{\epsfbox{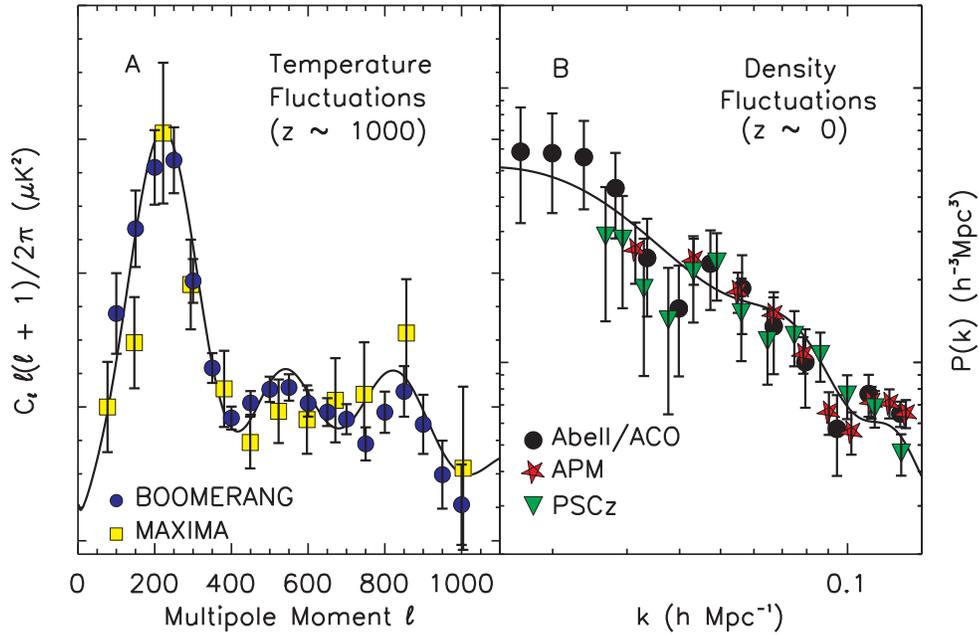}}
\caption{On the right we show the p(k) data from clusters and galaxies as
discussed by Miller et al. (2001a). The best fit model (with acoustic
oscillations) is shown. On the left is the 2001 CMB balloon data. Here we plot
the same model as shown on the right; {\it this is not a fit to the CMB
data}. As discussed in Miller et al. (2001b), there is good agreement between
the two epochs.  }
\end{figure}

In addition to the hunt for the turn--over in the p(k), several authors have
looked for ``bumps'' and ``wiggles'' in the local matter p(k). For example,
Einasto et al. (1997) reported a ``bump'' in the cluster p(k) which could be
do to excess power in the initial p(k) coming out of Inflation (a physical
mechanism for such a feature however, remains unclear; see Atrio--Barandela et
al. 2000). Alternatively, one may expect fluctuations in the observed p(k)
because of the acoustic oscillation of baryons in the early Universe;
sometimes called the ``Baryon Wiggles''.  Eisenstein et al. (1998) recently
looked for this signature in the p(k) measurements from galaxy and cluster
surveys but was unable to find a satisfactory fit between the cosmological
models (with acoustic oscillations) and the data available at that time. This
year, Miller et al. (2001a,b) and Percival et al. (2001) have revisited this
problem and have found compelling evidence for the detection of the ``Baryon
Wiggles'' in the local matter p(k) that is fully consistent with the detection
of the same ``Baryon Wiggles'' in the recent CMB data (MAXIMA. DASI \&
BOOMERANG). This concordance is shown in detail in Fig. 6.  In summary. in
2001, we have witnessed the simultaneous discovery of the ``Baryon Wiggles''
in the local ($z\sim0.1$) and distant ($z\simeq1000$) Universe providing yet
another major success for the Hot Big Bang cosmological model.

\section{The X--ray Cluster Luminosity Function}
\label{xc}

In addition to studying the clustering of clusters, it is important to
understand the distribution of cluster properties ({\it i.e.} the luminosity,
temperature and mass functions) and how these scale with each other and as a
function of redshift.  This will allow us to determine if clusters are a
self--similar population ({\it i.e.} their properties can be directly
predicted from using just the virial theorem and hydrostatic equilbrium, see
Kaiser 1986) or are there extra contributions to the energy which breaks the
simple scaling laws predicted by self-similarity.  In recent years, there have
been significant advances in understanding these scaling relationships (Horner
et al. 1999; Sheldon et al. 2001; Mathiesen 2001; Neumann \& Arnaud 2001) but
we can expect continued advancements over the next decade because of new
X--ray (XMM \& {\it Chandra}) and lensing (Joffre et al. 2000) surveys of
clusters as well as more detailed Hydro/N--body cluster simulations (see
Ricker \& Sarazin 2001). One example is the XMM-Newton $\Omega_m$ Project
(Bartlett et al. 2001) which is designed to obtain accurate temperature
measurements ($\simeq 10\%$ error) for all of the high redshift SHARC clusters
(Romer et al. 2000; Burke et al. 1997) thus obtaining an estimate of the high
redshift $L_{\rm x}$--$T_{\rm x}$ relation (over a decade in $L_{\rm x}$) as
well as constraining $\Omega_m$.

\begin{figure}[t]
\epsfxsize=3.8in 
\centerline{\epsfbox{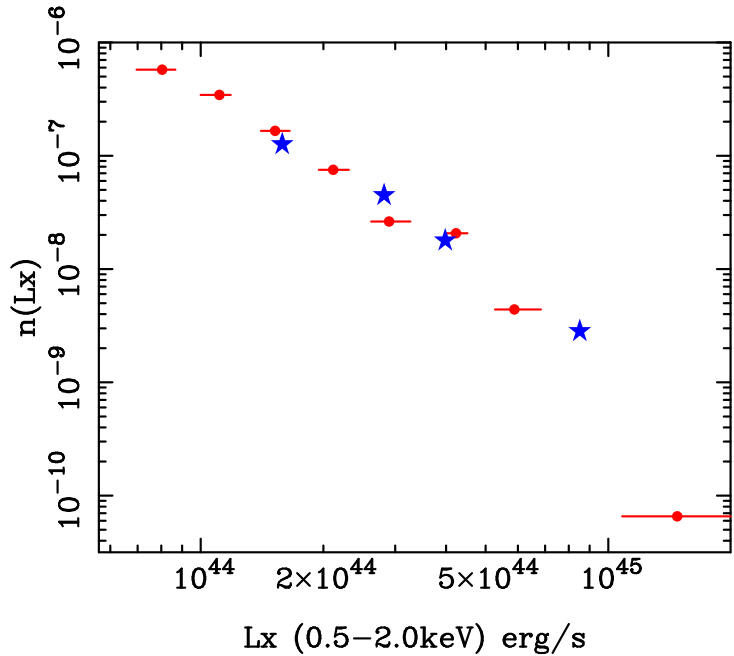}}
\caption{The X--ray Cluster Luminosity Function for the REFLEX (red;
Boehringer et al. 2001) and Bright SHARC (blue; Adami et al. 2000) cluster
surveys. The REFLEX probes to $z=0.3$, while the Bright SHARC surveys the
range $0.3<z\sim0.8$. The error bars on the REFLEX sample are discussed in
Boehringer et al. (2001), while a detailed analysis of the errors on the
Bright SHARC are given in Adami et al. (2000) and Romer et al. (2000). The
units on n(${\rm L_x}$) are $h^{-3}{\rm Mpc^{-3}\,(L_x(44))^{-1}}$. Data
kindly provided by Hans Bohringer and Christophe Adami.
\label{xclf}
}
\end{figure}

In the absence of a large, homogeneous sample of cluster temperatures, masses
and velocity dispersions, authors have focused in recent years on the X--ray
Cluster Luminosity Function (XCLF). The reader is referred to Gioia et
al. (2001) and Boehringer et al. (2001) for a good recent overview of ROSAT \&
{\it Einstein} measurements of the XCLF as a function of redshift. The key
debate with the XCLF is whether the bright--end (brighter than
$L^{\star}\simeq 4\times10^{44}\,{\rm erg/s}$) of the luminosity function has
evolved since $z\sim0.8$, thus producing a deficit of high redshift, luminous
systems. As discussed by many authors, the evolution of such massive clusters
is a strong constraint on $\Omega_m$ and thus it is vital to quantify how the
luminous end of the XCLF changes with look--back time ({\it e.g.}  Oukbir \&
Blanchard 1997; Viana \& Liddle 1999; Reichart et al. 1999; Borgani et
al. 1999b; Henry 2001).  If $\Omega_m\sim0.2$ ({\it e.g.} from p(k) of
clusters above), then the amount of evolution out to $z\simeq0.8$ should be
small and thus the signal we are looking for will be weak. In this case, we
must be very careful about systematic uncertainties in our distant cluster
surveys. Alternatively, if $\Omega_m=1$, we expect approximately an order of
magnitude decrease in the space density of $L_{\rm x}=10^{45}{\rm erg/s}$
systems to $z=0.8$.

In Fig. 7, I show the bright--end of the XCLF derived from the REFLEX cluster
survey (Boehringer et al. 2001) and the Bright SHARC (Romer et al. 2000; Adami
et al. 2000). These are two of the most well--understood surveys of X--ray
clusters available to date (in terms of their selection functions {\it
etc.}). The agreement between the two is remarkable but the number of luminous
clusters ($>L^{\star}$) in both surveys is still relatively small because of
the tiny space densities of such clusters. Furthermore, to obtain a ``high
precision'' measurement of the XCLF, we need to carefully model the selection
functions of these, and future, surveys.  This was recently emphasized by
Adami et al. (2000) who performed extensive simulations of the Bright SHARC
selection function and found that the shape and morphology of the clusters had
a dramatic effect on the efficiency of detection. When this is combined with
the effect seen in Fig. 1 -- {\it i.e.} that a large fraction of high redshift
clusters are likely to have signs of recent interactions (see also Henry 2001)
-- it becomes essential to model such effects in the selection function of any
distant cluster survey.  Therefore, the selection functions of future surveys
will need to include cosmological ({\it e.g.}  $\Omega_{\Lambda}$),
morphological ({\it e.g.} profile of clusters) and survey design ({\it e.g.}
off-axis PSF) effects.

\section{Future Cluster Surveys}

As emphasized above, future cluster surveys need to improve in both quality
and quantity. I highlight here four on--going efforts that will hopefully
address both of these points (this is by no means a complete list). First, the
MACS survey (Ebeling et al. 2001) is mining the ROSAT All--Sky Survey (RASS)
for previously undiscovered high redshift ($0.3<z<0.7$) luminous clusters. The
survey has been highly successful and has nearly 100 new massive clusters to
date.  Secondly, Gladders \& Yee (2000) are embarked on a large--area
($100{\rm deg^2}$), two--passband, optical survey in search of high redshift
clusters. Their technique makes use of the fact that cluster ellipticals are
strongly clustered in color and are the reddest (rest--frame) galaxies in
existence. They plan to expand their technique to $\simeq1000{\rm deg^2}$.

As discussed in Romer et al. (2001), the XMM data archive will be a gold--mine
for searching for distant X--ray clusters. Over the life--time of the
satellite, XMM is expected to serendipitously survey $\simeq800{\rm deg^2}$
which should yield thousands of cluster detections. Most importantly, the XMM
Cluster Survey (XCS; Romer et al. 2001) is predicted to find zero
($\Omega_m=1$), twelve ($\Omega_m=0.3$) or fifty ($\Omega_m=0.3\, + \,
\Omega_{\Lambda}=0.7$) massive clusters ($T_{\rm x}>6\,{\rm keV}$) at
redshifts greater than one (see Romer et al. 2001).  This will provide an
excellent constraint on both $\Omega_m$ and $\Omega_{\Lambda}$ as well as
providing a rich dataset of cluster temperatures and morphologies.

Finally, the Sloan Digital Sky Survey (SDSS) should produce several high
quality, large--area, cluster surveys. For example, in Nichol et al. (2000), I
outlined a new algorithm called ``C4'' which uses the 5 passband SDSS
photometry and photometric redshifts, to look for clusters in a 7--dimensional
space (4 colors, 2 angular, 1 radial). This removes all concerns about
projection effects (see Section 2) and, when combined with the RASS data, can
find clusters out to $z\simeq0.7$.

Before I leave this section, I will emphasize again the importance of
determining the selection functions of these new surveys.  As the data,
procedures and algorithms used in these new surveys get more complex, then we
can not naively assume the volume of these surveys is just calculated from
their flux limit and their area on the sky. We will need to turn to
simulations of the Universe to help us {\it e.g.} using semi--analytical,
N--body techniques (see Kauffmann et al. 1999) or fast analytical
approximations like PTHALO (Scoccimarro \& Sheth 2001). I note here that
Postman (this volume) also emphasizes the need for well--understood selection
functions for distant cluster surveys.

I thank the organizers of the conference for inviting me.  I would like to
also thank my collaborators on the SDSS, XMM--Newton $\Omega_m$ project, XCS
and SHARC surveys. I especially thank Chris Miller, Chris Collins and Kathy
Romer for their help, guidance and reading of this manuscript.

\section{References}
\noindent Abadi, M., et al. 1998, ApJ, 507, 526\\ 
Adami, C., et al. \ 2000, \apjs, 131, 391 \\
Atrio-Barandela, F., et al. 2000, ApJL, see astro-ph/0012320\\
Bahcall, N.~A.~\& Soneira, R.~M.\ 1983, \apj, 270, 20 \\
Bartlett, J., et al. 2001, see astro-ph/0106098\\
Biviano, A., 2000, see stro-ph/0010409\\
Boehringer, H., et al. 2001, see astro-ph/0106243\\
Borgani, S., Plionis, M., \& Kolokotronis, V.\ 1999a, \mnras, 305, 866 \\
Borgani, S., et al.\ 1999b, \apj, 527, 561 \\
Borgani, S.~\& Guzzo, L.\ 2001, Nature, 409, 39 \\
Burke, D.~J., et al. \ 1997, \apjl, 488, L83\\ 
Chiu, W.~A., Ostriker, J.~P., \& Strauss, M.~A.\ 1998, \apj, 494, 479 \\
Cohn, J.~D., Bagla, J.~S., \& White, M.\ 2001, \mnras, 325, 1053 \\
Collins, C.~A.~et al.\ 2000, \mnras, 319, 939 \\
Croft, R.~A.~C., et al. \ 1997, \mnras, 291, 305 \\
Dalton, G.~B. et al.\ 1992, \apjl, 390, L1 \\
Dekel, A., Blumenthal, G.~R., Primack, J.~R., \& Olivier, S.\ 1989, \apjl, 
338, L5 \\
Ebeling, H., Edge, A.~C., \& Henry, J.~P.\ 2001, \apj, 553, 668 \\
Efstathiou, G. et al.\ 1992, \mnras, 257, 125 \\
Einasto, J., et al. 1997, Nature, 385, 139\\
Eisenstein, D.~J., Hu, W., Silk, J., \& Szalay, A.~S.\ 1998, \apjl, 494, L1 \\
Eisenstein, D.~J., et al. see astro-ph/0108153\\
Gal, R.~R., et al. \ 2000, \aj, 120, 540\\ 
Gioia, I.~M., et al.\ 2001, \apjl, 553, L105 \\
Gladders, M.~D.~\& Yee, H.~K.~C.\ 2000, \aj, 120, 2148 \\
Guzzo, L., 2001, see astro-ph/0102062 \\
Haiman, Z., et al. 2001, see astro-ph/0103049\\
Henry, J. P., 2001, astro-ph/0109498\\
Horner, D.~J., Mushotzky, R.~F., \& Scharf, C.~A.\ 1999, \apj, 520, 78 \\
Huchra, J.~P., Geller, M.~J., Henry, J.~P., \& Postman, M.\ 1990, \apj, 
365, 66 \\
Huterer, D., Knox, L., \& Nichol, R.~C.\ 2001, \apj, 555, 547 \\
Jenkins, A., et al. 2001, \mnras, 321, 372 \\
Joffre, M.~et al.\ 2000, \apjl, 534, L131 \\
Kaiser, N.\ 1986, \mnras, 222, 323 \\
Kauffmann, G. et al.\ 1999, \mnras, 303, 188\\ 
Kerscher, M.~et al.\ 2001, \aap, 377, 1\\
Lahav, O., Fabian, A.~C., Edge, A.~C., \& Putney, A.\ 1989, \mnras, 238, 881 \\
Ling, E., Frenk, C., Barrow, J., 1986, MNRAS, 223, 21\\
Lumsden, S.~L. et al.\ 1992, \mnras, 258, 1 \\
Matarrese, S., Verde, L., \& Jimenez, R.\ 2000, \apj, 541, 10 \\
Mathiesen, B.~F.\ 2001, \mnras, 326, L1\\
Mathiesen, B.~F.~\& Evrard, A.~E.\ 2001, \apj, 546, 100 \\
Miller, C.~J., Batuski, D.~J., Slinglend, K.~A., \& Hill, J.~M.\ 1999, 
\apj, 523, 492 \\
Miller, C.~J., Nichol, R.~C., \& Batuski, D.~J.\ 2001a, \apj, 555, 68\\
Miller, C.~J., Nichol, R.~C., \& Batuski, D.~J.\ 2001b, Science, 292, 2302\\
Moscardini, L. et al.\ 2000, \mnras, 314, 647 \\
Narayanan, V.~K., Berlind, A.~A., \& Weinberg, D.~H.\ 2000, \apj, 528, 1 \\
Neumann, D.~M.~\& Arnaud, M.\ 2001, \aap, 373, L33 \\
Nichol, R.~C., Briel, O.~G., \& Henry, J.~P.\ 1994, \mnras, 267, 771 \\
Nichol, R.~C. et al.\ 1992, \mnras, 255, 21P\\ 
Nichol, R. C., et al. 2000, astro-ph/0011557\\
Oukbir, J.~\& Blanchard, A.\ 1997, \aap, 317, 1 \\
Peacock, J., \& West, M., MNRAS, 259, 494\\
Percival, W., 2001, MNRAS, see astro-ph/0105252\\
Postman, M., Huchra, J.~P., \& Geller, M.~J.\ 1992, \apj, 384, 404 \\
Reichart, D.~E., et al. \ 1999, \apj, 518, 521\\ 
Ricker, P., Sarazin, C., 2001, ApJ, see astro-ph/0107210\\
Ritchie, B. W., Thomas, P. E., MNRAS, see astro-ph/0107374\\
Robinson, J., Gawiser, E., \& Silk, J.\ 2000, \apj, 532, 1 \\
Roettiger, K., Burns, J.~O., \& Loken, C.\ 1996, \apj, 473, 651 \\
Romer, A.~K.~et al.\ 1994, Nature \\
Romer, A.~K.~et al.\ 2000, \apjs, 126, 209 \\
Romer, A.~K.~et al.\ 2001, \apj, 547, 594 \\
Schindler, S., 2001a, see astro-ph/0109040\\
Schindler, S., 2001b, see astro-ph/0107008\\
Schuecker, P.~et al.\ 2001, \aap, 368, 86 \\
Scoccimarro, R., Sheth, R. K., 2001, see astro-ph/0106120\\
Sheldon, E.~S.~et al.\ 2001, \apj, 554, 881\\
Sutherland, W.\ 1988, \mnras, 234, 159 \\
Szalay, A. S., 2001, ApJ, see astro-ph/0107419\\
Tegmark, M.\ 1999, \apjl, 514, L69 \\
Verde, L. et al. \ 2001, \mnras, 325, 412 \\
Viana, P.~T.~P.~\& Liddle, A.~R.\ 1999, \mnras, 303, 535 \\
West, M., \& van der Bergh, S., 1991, 373, 1\\
White, S.~D.~M., Frenk, C.~S., Davis, M., \& Efstathiou, G.\ 1987, \apj, 
313, 505 \\

\end{document}